\documentclass{evn2004}
\usepackage{txfonts}
\usepackage{graphicx}
\begin{document}
\setcounter{page}{297}

   \title{Analysis strategies and software for geodetic VLBI}

   \author{R. Haas}

   \institute{Onsala Space Observatory,
Chalmers University of Technology, 
SE-439 92 Onsala, Sweden}

   \abstract{
This article describes
currently used analysis strategy and 
data analysis software for geodetic VLBI.
Today's geodetic observing strategies are shortly presented,
and the geodetic VLBI observables and
data modeling are briefly discussed.
A short overview is given on existing geodetic VLBI
software packages and the statistical approaches that are applied.
Necessary improvements of today's analysis software
are described.
Some of the future expectations and goals of
geodetic VLBI are presented and the
corresponding consequences for the VLBI technique are explained.
This includes consequences in terms of technical
development and corresponding improvements in data modeling and 
analysis software.
   }

   \maketitle

%________________________________________________________________

\section{Introduction}

Since 1999 the International VLBI Service for Geodesy and Astrometry 
coordinates and schedules global geodetic VLBI sessions
(Schl{\"u}ter et al. \cite{schluter02}).
It applies observing strategies that aim at exploiting the available
resources for geodetic VLBI in a best possible way in order
to reach today's research goals.
These observing strategies are briefly described in Section~2.

The IVS analysis centers apply analysis strategies and 
use analysis software that allow to analyze today's
geodetic VLBI observables as good as possible. 
This is shortly described in sections 3 to 6.
Using these approaches important scientific results 
can be derived from the geodetic VLBI data analysis.

However, further improvements for example in modeling 
of geophysical effects and error sources are necessary 
in order to gain further insights in today's research achievements.
These are described briefly in Section 7.

Furthermore, in order to reach new research goals, 
technical improvements of the geodetic VLBI technique
and the corresponding improvements in data modeling, 
analysis software and analysis strategy are necessary.
They are described in sections 8 and 9.

%__________________________________________________________________

\section{Geodetic VLBI observing strategies}

The IVS coordinates and schedules global geodetic VLBI 
observing sessions.
This is done in agreement with the contributing partner
organizations, i.e. the international radio astronomical observatories 
and the correlators that are active in geodetic VLBI. 
The main objectives of the coordinated observation
sessions are contributions to reference systems and 
their relations, 
i.e. the International Terrestrial Reference Frame
(ITRF, e.g. Altamimi et al. \cite{altamimi02}) 
the International Celestial Reference Frame 
(ICRF, e.g. Ma et al. \cite{ma98}), 
and time series of Earth Orientation Parameters (EOP).
Currently, dedicated observation sessions are scheduled that are
optimized for the different goals. The available resources in terms of
radio telescopes and correlator capacity act of course as constraints.

The IVS schedules multi-station sessions for 
EOP determination twice a week.
Here large and distributed networks of currently up to 
eight stations perform observations for 24 hours.
The full set of EOP components, 
i.e. corrections to the nutation model,
polar motion components and the earth rotation component UT1,
can be determined 
from the data analysis of these observation sessions.

Daily short sessions are scheduled to observe at least
the earth rotation component every weekday. 
For these sessions only two stations that form an
extended east-west baseline observe for two hours.

Multi-station observation sessions for the ITRF and the ICRF are
scheduled several times per year with large and distributed
networks and observing time of 24 hours.

These observing strategies exploit today's existing resources
for geodetic VLBI in a nearly optimum way.
However, from a scientific point of view it would be desirable
to use a more holistic approach and to optimize the observations
sessions in a way that  all parameters of interest can be 
determined simultaneously. 
This means that the geodetic VLBI technique has
to be developed further and for example the number of
observing instruments has to be increased
(Petrachenko et al. \cite{petrachenko04}).

%__________________________________________________________________

\section{The geodetic VLBI observables}

A successful correlation of geodetic VLBI observations 
results in phase delay and group delay observables.
The use of the phase delays for geodetic purposes is
complicated because of loss of phase coherence between 
different scans due to instrumental and environmental
influences (Campbell \cite{campbell00}). 
Geodetic VLBI observation schedules are usually optimized 
for sky coverage, i.e. observing the widest possible 
distribution of radio sources over the sky per time interval. 
This scheduling strategy is applied in order to be able 
to successfully determine the geodetic parameters of 
interest in these time intervals. 
The drawback is that the telescopes move  
over large slewing ranges and phase coherence 
between succeeding scans is lost.

So far it has not been completely successful to use phase
delay in geodesy (Herring \cite{herring92}, Petrov \cite{petrov99}).
Thus, only the group delay observables are analyzed routinely. 
The precision of the group delay observables is today 
on the level of 10~ps (Sovers \& Fanselow \cite{sovers98}).
This precision is currently not high enough to reach the goal of
long-term accuracy of geodetic reference frames at the level
of 1~mm or below.
Therefore the geodetic VLBI technique needs further development
that allows to increase the delay precision
(Petrachenko et al. \cite{petrachenko04}).

%__________________________________________________________________

\section{Data modeling}

Any analysis of observed data of course requires 
also data modeling which is based on theoretical 
models and a~priori information.
The modeling of so-called theoretical observations
and the application of statistical analysis methods 
form the basis to derive the parameters of interest.
For geodetic VLBI this means that the group delay 
observables have to be modeled accurately
(e.g. Sovers \& Fanselow \cite{sovers98}).

The geometric delay is the largest component of the 
observed group delay.
It is modeled in a quasi-inertial solar system 
barycentric (SSB) frame based on a~priori 
information of the proper station locations 
and the direction to the observed radio source.

Before the geometric delay in the SSB can be calculated,
the proper locations of the stations given in the
earth fixed frame have to be transformed to the SSB frame.
Station displacements due to plate tectonics, 
solid earth tides, pole tide, 
ocean tide loading, and atmospheric 
loading need to be modeled.
The two frames involved have to be aligned using the state of the art 
precession-nutation model and a~priori information 
about polar motion and earth rotation and the 
corresponding tidal variations of the EOPs.
The transformation from a geocentric to a barycentric
frame is performed by a Lorentz transformation and includes
relativistic effects.
Then the geometric delay can be calculated and corrected
for general relativistic effects and transformed to a 
proper delay.
This delay is transformed back to a geocentric frame
via a Lorentz transformation.

The propagation of the radio waves through the 
earth's atmosphere also has to be considered.
The ionospheric contribution is usually corrected for
based on the dual-frequency observations at S- and X-band.
The tropospheric delay is usually modeled
a~priori based on local pressure observed
at the stations and the corresponding mapping functions.
Today still widely used mapping functions 
use season, latitude and altitude as
input parameters (Niell \cite{niell96}).

More details on data modeling can for example be found in 
Sovers \& Fanselow ({\cite{sovers96})
and Sovers et al. ({\cite{sovers98}).
The data modeling as described above should in general 
follow the recommendations of the International Earth
Rotation and Reference Systems Service (IERS) that are 
formulated as so-called IERS Conventions.
These form a common and consistent basis 
for all different geodetic space techniques to 
allow comparisons and combinations
of the results derived from these 
techniques in a meaningful way.

%__________________________________________________________________

\section{Analysis strategies}

It is common use in geodetic VLBI analysis, and also in other
geodetic space techniques, to distinguish between so-called
arc parameters and global parameters.
This distinction reflects the time
epochs for which the parameters are valid.

Arc parameters are parameters that are valid only during 
a particular observation session or parts of it.
Examples are parameters that describe the turbulent troposphere,
relative clock parameters, and the earth orientation parameters 
that relate the earth-fixed and quasi-inertial reference frames. 

Global parameters are  on the other hand valid for longer 
time period and not only for the actual observing session. 
For example radio source coordinates, relativistic parameters,
and station coordinates and
velocities belong to this category of parameters.

A so-called single-session analysis uses only VLBI
observables of one session of usually 24 hours
duration.
With this approach only arc parameters can be 
accurately determined.
This means that for example the station coordinates and radio
source positions are kept fixed at their a~priori values and
the EOP's are determined from the data analysis.
Of course also tropospheric and clock parameters have
to be estimated.
With this strategy it is possible to derive EOP for individual
observing sessions and thus finally a time series of EOP. 

It is also possible to derive time series of relative 
station positions using the single-session approach. 
In this case the a~priori EOP and radio source positions
are kept fixed and relative station coordinates and tropospheric
and clock parameters are determined.
These relative station coordinates are of course valid only 
for the epoch of the observation session.
In a second analysis step using the time series
of relative coordinates, also relative station velocities
can be determined (e.g. Haas et al. \cite{haas03}).

A so-called global analysis uses a large number of VLBI 
sessions together and allows to solve for both arc and 
global parameters.
One possible approach is to accumulate reduced normal 
equations from single sessions that no longer contain arc parameters,
and then to solve for the global parameters.
After that the arc parameters for each session can be determined
in a second step by substituting the estimated global 
parameters.
Another possible approach is to combine the 
variance-covariance matrices of 
individual observing sessions with filtering techniques 
and to determine stochastic parameters 
in a smoothing approach (Andersen \cite{andersen00}).

Some interesting results from global analysis are
for example investigations of the free core nutation 
(Herring et al. \cite{herring86}),
tidal effects in the earth rotation (Brosche et al. \cite{brosche91}),
ocean tide loading (Sovers \cite{sovers94}),
frequency dependent  Love and Shida numbers 
(Haas \& Schuh \cite{haas96}),
the ICRF (Ma et al. \cite{ma98}),
atmospheric loading (Petrov \& Boy \cite{petrov04}), and
general relativity (Shapiro et al. \cite{shapiro04}).

%__________________________________________________________________

\section{Data analysis software}

A number of geodetic VLBI data software packages 
have been developed during the last decades.
Table~1 is an attempt to give an overview of these
software packages, though not claiming completeness.
Unfortunately, not all software packages are documented 
in an easy accessible way. 
Some of the packages have stopped maintainance and development,
e.g. VORIN,  others are still in a process of 
development, e.g. QUASAR.
It appears that currently the VLBEST software package is the only
one to allows automated real-time data analysis, 
as demonstrated in the
Japanese Keystone project (Koyama et al. \cite{koyama98}).
The two software packages GEOSAT and GINS were
developed with the aim to integrate VLBI data analysis
with data analysis of other geodetic space techniques
at the observational level.

%__________________________________________________
%__________________________________________________
\begin{table}[b!]
\caption{Geodetic VLBI data analysis software packages.
}
\begin{center}
\begin{tabular}{ll}
\hline
\noalign{\smallskip}
Software package & Statistical Method$^{*}$ \\
\noalign{\smallskip}
\hline
CALC/SOLVE (Ma et al. \cite{ma90}) & LSQ\\
OCCAM (Titov et al. \cite{titov01}, \cite{titov04})  & LSQ/KF/LSQC\\
MODEST (Sovers \& Jacobs \cite{sovers96}) & SRIF\\
SOLVK (Herring et al. \cite{herring90}) & KF \\
STEELE-BREEZE (Bolotin \cite{bolotin00}) & SRIF \\
GLORIA (Gontier, \cite{gontier92}) & LSQ \\
VLBEST (Koyama et al. \cite{koyama98}, \cite{koyama99}) & LSQ \\
GEOSAT (Andersen \cite{andersen95}, \cite{andersen00}) & KF\\
VORIN (Petrov \cite{petrov95}) & LSQ \\
ERA (Krasinsky \& Vasyliev \cite{krasinsky97}) & LSQ\\
GINS (Meyer et al. \cite{meyer00}) & LSQ\\
QUASAR (Gubanov et al. \cite{gubanov04}) & LSQC\\
\noalign{\smallskip}
\hline
\noalign{\smallskip}
\end{tabular}
\end{center}
{*The abbreviations used for the statistical methods are:\\
LSQ -- Least-Squares method,
SRIF -- Square-Root Information Filter,
KF -- Kalman Filter,
LSQC -- Least-Squares Collocation method.
See text for further explanation.}
\end{table}
%__________________________________________________
%__________________________________________________

In general, the software packages aim at following
some general modeling advice for geodetic
space techniques that has been agreed on 
in the geodetic community and is formulated 
in the IERS Conventions.
Most software packages claim to comply with the
IERS Conventions 1996 (IERS \cite{iers96}), 
and several are on the way or have 
already updated to the 
IERS Conventions 2003 (IERS \cite{iers03}).

Some inter-comparison tests between individual
software packages have been performed in the past
(e.g. Sovers \& Ma \cite{sovers85}).
However, so far there has not been a common
comparison between all software packages.

The software packages use various statistical 
methods for the actual data analysis. 
These statistical methods include 
the Least-Squares (LSQ) method (e.g., Koch \cite{koch88}),
the Kalman-filter (KF) method (e.g. Kalman \cite{kalman60}),
the Square-Root Information Filter (SRIF)
(Bierman \cite{bierman77})
and the Least-Squares Collocation (LSQC) method
(e.g. Koch \cite{koch88}, Moritz \cite{moritz00}).
These statistical approaches differ mainly in 
the way the variance-covariance information 
is propagated and the ability to treat 
stochastic parameters.

Most of the software packages aim at 
allowing portability to a large number of 
computer platforms and operating systems.
However, this goal has so far only been
reached for very few cases and still some
dependency on computer platform and 
operating system exists.
For example the widely used software package
SOLVE is still only available for HP machines
and HP-Unix operating systems.

The software packages still use different data input formats.
Both binary and ASCII data formats are in use and
the necessary conversion software exists.
However, a working group of the IVS
tries to establish a common exchange format
called PIVEX (Gontier \& Feissel \cite{gontier02}).

Currently there are 7 full IVS analysis centers
and 14 associated analysis centers.
The CALC/SOLVE software is used by 9,
OCCAM by 4, and MODEST by 2,
while GLORIA, GEOSAT, VLBEST, SOLVK, and STEELE-BREEZE
are each used by one analysis center only.

%______________________________________________________________

\section{Necessary improvements today}

Although important results can be derived from
today's geodetic VLBI data analysis, 
further improvements in particular in the fields
of data modeling and statistical methods
are necessary.
These improvements are needed today and 
independent of possible technical modifications
of the geodetic VLBI technique that might lead 
to higher precision of the observables.

Some of the current limitations on the modeling side
are due to insufficient atmospheric modeling.
Mapping functions based on Numerical Weather 
Models (NWM) promise to lead to improvements
(Boehm and Schuh \cite{boehm04}, 
Stoyanov et al. \cite{stoyanov04}).
An even more interesting approach might be to 
apply direct raytracing through NWM instead of 
using mapping functions.
Also modeling based on turbulence theory
appears to be an interesting approach
(Emardson \& Jarlemark \cite{emardson99}).
Thus, the data analysis packages should be 
extended to incorporate these approaches.

Another concern of improved data modeling 
is radio source structure.
In the ideal case the radio sources observed
for geodetic VLBI would all be structureless 
compact objects, i.e. point sources. 
However, there are many sources that show considerable structure
at X-band (Fey \& Charlot \cite{fey98}, \cite{fey00}).
Thus, there is a need to model the source structure effects
and to incorporate this in the geodetic VLBI data analysis.
So far this is not done on a regular basis and therefore
the software packages have to be extended to be able to do so.
This is true even for the case that future geodetic VLBI 
observations might use higher frequencies with less structure,
since it has to be guaranteed that the historic observations 
can be re-analyzed in the best possible way.

Periodic station displacements due to solid earth tide and
ocean tide loading effects are modeled routinely
and with high precision in today's data analysis.
However, non-periodic station displacements due to
atmospheric and hydrological loading or local deformation
of the telescopes as a function of temperature, are not yet
treated routinely.
It appears that atmospheric loading can be modeled with
sufficient accuracy based on convolution of global pressure data
(Scherneck et al. \cite{scherneck02},
Petrov \& Boy \cite{petrov04}).
Thus, the modeling of this phenomenon should be incorporated in all data
analysis software packages. 
Hydrological loading is more difficult to
model mainly because the hydrological models are still restricted in
accuracy. 
Therefore this loading effect will still be a topic 
of investigation for the future.

Thermal deformation of radio telescopes is today monitored
routinely at two of the radio telescopes used for geodetic
VLBI observations.
A simple model to describe the thermal deformation effect
is presented in the IERS Conventions 2003 (IERS \cite{iers03}).
However, so far neither the actual deformation measurements
nor the model is used routinely in all data analysis packages.
More advanced modeling based on finite element calculations
promises to allow modeling for any kind of telescope
(Clark \& Thomsen \cite{clark88}) and might be incorporated
in the data analysis in the near future.

Currently there are also limitations in the 
statistical part of the data analysis.
The existing data analysis packages differ concerning the
statistical methods that are applied and how stochastic
parameters are treated.
It appears that in some cases there are deficiencies
in particular in the handling of covariances between 
different parameters (Tesmer \& Kutterer \cite{tesmer04}). 
Thus, the software packages should be extended and more
refined stochastic models should be incorporated.

%______________________________________________________________

\section{New scientific goals}

The near future goals of geodetic VLBI are to achieve a 
long-term accuracy of geodetic reference frames on the 
1~mm level or better. 
In this contexts the consistency of the reference frames and the EOP is 
of major concern and requires rigorous analysis 
solutions (Schuh et al. \cite{schuh04}). 
A holistic approach for the planning of observation 
sessions and the corresponding rigorous data analysis 
is desirable.
However, it will require further development of the geodetic VLBI 
technique as such, and the establishment of additional
radio telescopes (Petrachenko et al. \cite{petrachenko04}). 

One goal concerning the terrestrial reference frame is
for example an improved treatment of periodic and
aperiodic effects in order to achieve a more
robust reference frame.
This is related to the question of
geodynamical modeling.
For the celestial reference frame one goal is to 
densify the radio source catalogue and
to observe also weaker sources.
Of particular interest is the connection between the
quasi-inertial reference frames and the
dynamical reference frames.

Besides pure reference frame investigation, other goals are
to intensify the investigation of a number of 
geodynamical phenomena.
Among these are for example processes in the 
earth interior that are related to Free Core 
Nutation (FCN) and Free Inner Core Nutation (FICN).
The investigations will require improvements 
in the data modeling in order to be able to
increase sensitivity for these phenomena.
Also the question of the earth's free oscillations 
is of increasing importance in the geodynamical
context. 

A better understanding of the governing 
geodynamical mechanisms that cause 
EOP variations in the sub-diurnal
frequency band is another important research topic.
One approach to this research is to resolve
high-frequent EOP from continuous VLBI 
observations with large and
geometrically well distributed networks.

Further information on new goals for geodetic VLBI
can be found for example in Schuh et al. (\cite{schuh04}). 

%______________________________________________________________

\section{Necessary future developments}

Further development of the geodetic VLBI technique
is necessary in order to be able to address the new 
scientific goals and to contribute to improvements of
today's scientific achievements.
This development has to fight current limitations
in technology, data modeling, and data analysis.

One technical limitation of today's geodetic VLBI
observations and data analysis is the increasing
amount of radio frequency interference (RFI)
caused by communication operators.
Both satellite based and ground based communication links
disturb in particular the S-band observations and endanger
the possibility to compensate for ionospheric
effects with the current S/X-frequency set-up used
in geodetic VLBI.
Thus, there are considerations in the IVS to
modify the geodetic frequency set-up
(Petrachenko et al. \cite{petrachenko04}).
Observations is the S/X bands will have to be continued
in order to guarantee continuity for the existing ICRF,
but both, lower and higher observing frequencies could be added.

Higher frequency observations in the K-band could 
effectively avoid the interference problems.
Another advantage of higher frequency observations
is that the radio sources at these frequencies
appear to have less source structure 
(Boboltz et al. \cite{boboltz04}).

Lower frequencies observations in the L-band could allow also 
to observe signals of Global Navigation Satellite Systems (GNSS) and
in that way contribute to a combination of geodetic space techniques
and the integration of quasi-inertial and dynamical reference frames.

Observations at frequencies near the water vapor absorption 
line might allow using the VLBI telescopes directly
as line-of-sight water vapor radiometers.
These measurements could be used to compensate directly for 
tropospheric propagation effects instead of using
other external information 
(Petrachenko et al. \cite{petrachenko04}).

The possible change in the frequency set-up will
require development of the geodetic
VLBI hardware.
It also has to be reflected in the 
analysis software packages. 
More details on plans for a modified frequency set-up for 
geodetic VLBI can be found in
Petrachenko et al. (\cite{petrachenko04}).

Another more or less technical limitation is the
described loss of phase coherence that makes
it impossible to use phase delay observables.
Proposals to solve this problem aim also at
observing at more than two frequencies simultaneously
and at using a pair of telescopes at each site
(Petrachenko et al. \cite{petrachenko04}).
A technical development according to these ideas
will of course require developing corresponding
analysis strategies and to extend the 
existing analysis software.

A technical development that is currently ongoing
is intercontinental real-time e-VLBI.
Real-time observations of for example EOP are
interesting for reference frame research
and applications for navigation.
It is anticipated that such real-time
observations can be performed on a regular basis in the near future.
Thus the capability to perform automated analysis in
real-time should be added to all the existing 
data analysis software packages.

%______________________________________________________________

\section{Conclusions}

Today's analysis strategy and data analysis software for geodetic
VLBI correspond to current observing strategy and accuracy of the
VLBI group delay observables.
This set-up exploits today's resources in a nearly optimum way.
Interesting and important geophysical and geodynamical
results can be derived from geodetic VLBI data analysis.
However, some improvements in the fields of data modeling and 
statistical methods are necessary even for today's observations.

In order to live up to the future scientific expectations 
and in order to address new scientific goals in geodetic VLBI,
further development is necessary.
Technical development is required in order to a reach higher precision
of the VLBI observables.
The analysis strategy will have to correspond to possible changes in
observing strategies and for example concentrate primarily on global
analysis in a holistic set-up of observing sessions.
The analysis software packages need to be developed further
and improvement in data modeling and statistical methods
have to be incorporated.
Also the general ability to perform automated analysis 
in real-time, and to analyze additional observing frequencies
has to be added.

%______________________________________________________________

%\begin{acknowledgements}
%\end{acknowledgements}

%______________________________________________________________

%_____________________________________________________________
\cleardoublepage
%_____________________________________________________________


\begin{thebibliography}{}

 \bibitem[2002]{altamimi02}
  Altamimi, Z., Sillard, P. \& Boucher, C. 2002,
  J. Geophys. Res., 107(B10)

 \bibitem[1995]{andersen95}
  Andersen, P. H. 1995,
  NDRE Publ. 95/01094

 \bibitem[2000]{andersen00}
  Andersen, P. H. 2000,
  J. Geodesy, 74(7--8), 531--551

 \bibitem[1977]{bierman77}
  Bierman, G. 1977,
  Factorization Methods for Discrete Sequential Estimation, 
  (Academic, New York)

 \bibitem[2004]{boboltz04}
  Boboltz, D. A., Fey, A. L., Charlot, P.,
  Fomalont, E. B., Lanyi, G. E., Zhang, L. D.
  \& KQ VLBI Survey Collaboration 1004,
  in IVS 2004 General Meeting Proceedings, 
  ed. by N. R. Vandenberg \& K. D. Baver, 361--365

 \bibitem[2004]{boehm04}
  Boehm, J. \& Schuh, H. 2004,
  Geophys. Res. Lett., 31

 \bibitem[2000]{bolotin00}
  Bolotin, S. 2000, SteelBreeze home page,
  available at http://steelbreeze.sourceforge.net

 \bibitem[1991]{brosche91}
  Brosche, P. W{\"u}nsch, J., Campbell, J. \& Schuh, H. 1991,
  A \& A 245, 676--682

 \bibitem[2000]{campbell00}
  Campbell, J. 2000,
  in IVS 2000 General Meeting Proceedings, 
  ed. by N. R. Vandenberg \& K. D. Baver, 19--34

 \bibitem[1988]{clark88}
  Clark, T. A. \& Thomsen 1988,
  NASA Technical Memorandum 100696

 \bibitem[1999]{emardson99}
  Emardson, T. R. \& Jarlemark, P. O. J. 1999,
  J. Geodesy 73(6), 322-331

 \bibitem[1998]{fey98}
  Fey, A. L. \& Charlot, P. 1998,
  ApJS 111, 95

 \bibitem[2000]{fey00}
  Fey, A. L. \& Charlot, P. 2000,
  ApJS 128, 17

 \bibitem[1992]{gontier92}
  Gontier, A.-M. 1992,
  These de doctorat de l'Observatoire de Paris

 \bibitem[2000]{gontier02}
  Gontier, A.-M. \& Feissel, M. 2000,
  in IVS 2002 General Meeting Proceedings, 
  ed. by N. R. Vandenberg \& K. D. Baver, 248--254

 \bibitem[2004]{gubanov04}
  Gubanov, V. S., Rusinov, Y. L, Surkis, I. F.,
  Kurdubov, S. L. \& Shabun, C. Y. 2004,
  in IVS 2004 General Meeting Proceedings, 
  ed. by N. R. Vandenberg \& K. D. Baver, 315--319

 \bibitem[1996]{haas96}
  Haas, R. \& Schuh, H. 1996,
  Geophys. Res. Lett. 23, 1509--1512

 \bibitem[2003]{haas03}
  Haas, R., Nothnagel, A., Campbell, J. \& Gueguen, E. 2003,
  J.~Geodyn., 35(4--5), 391--414

 \bibitem[1986]{herring86}
  Herring, T., A., Gwinn, C. R., , \& Shapiro I. I. 1986,
  J.~Geophys. Res., 91, 4745--4754

 \bibitem[1990]{herring90}
  Herring, T., A., Davis, J. L. \& Shapiro I. I. 1990,
  J.~Geophys. Res., 95(B8), 12561--12581

 \bibitem[1992]{herring92}
  Herring, T. A. 1992,
  J. Geophys. Res., 97, 1981--1990

 \bibitem[1996]{iers96}
  International Earth Rotation Service 1996,
  IERS Conventions (1996), 
  IERS Technical Note 21, ed. by D. D. McCarthy

 \bibitem[2003]{iers03}
  International Earth Rotation Service 2003,
  IERS Conventions (2003), 
  IERS Technical Note 32, ed. by D. D. McCarthy \& G. Petit
 
 \bibitem[1960]{kalman60}
  Kalman, R. E. 1960,
  J. Basic Engng., 95--108

 \bibitem[1988]{koch88}
  Koch, K. R. 1988,
  Parameter estimation and hypothesis testing in linear models
  (Springer, New-York)

 \bibitem[1998]{koyama98}
  Koyama, Y., Kurihara, N., Kondo, T., Sekido, M.,
  Takahashi, Y., Kiuchi, H. \& Heki, K. 1998,
  Earth Planets Space 50, 709--722

 \bibitem[1999]{koyama99}
  Koyama, Y., Heki, K., Takahashi, Y. \& Furuya, M. 1999,   
  Journal of the Communications Research Laboratory,
  46(1), 77--81

 \bibitem[1997]{krasinsky97}
  Krasinsky, G. A. \& Vasyliev, M. 1997,
  in Proc. IAA Coll. 165, Kluwer Acad. Publ., 239--244

 \bibitem[1998]{ma98}
  Ma, C., Arias, E. F., Eubanks, T. M., Fey, A. L.,
  Gontier, A.-M., Jacobs, C. S., Sovers, O. J.,
  Archinal, B. A., \& Charlot,~P. 1998,
  A. J. 116, 516--546

 \bibitem[1990]{ma90}
  Ma, C., Sauber, J. M., Bell, L. J., Clark, T. A.,
  Gordon, D. \& Himwich, W. E. 1990,
  J. Geophys. Res., 95, 21991--22011

 \bibitem[2000]{meyer00}
  Meyer, U., Charlot, P. \& Biancale, R. 2000,
  in IVS 2000 General Meeting Proceedings, 
  ed. by N. R. Vandenberg \& K. D. Baver, 324--328

 \bibitem[2000]{moritz00}
  Moritz, H. 2000,
  Mathematische Geologie, 5, 205--213

 \bibitem[1996]{niell96}
  Niell, A. E. 1996,
  J. Geophys. Res. 101, 3227--3246

 \bibitem[2004]{petrachenko04}
  Petrachenko, B., Corey, B., Himwich, E., Ma, C.,
  Malkin, Z., Niell, A., Shaffer, D. \& Vandenberg, N. 2004,
  Report of IVS WG3.1 - Observing strategies,
  available at http://ivscc.gsfc.nasa.gov/about/wg/wg3/index.html

 \bibitem[1995]{petrov95}
  Petrov, L. Y. 1995,
  Communications of the Institute of Applied
  Astronomy N74,75,76, Institute of Applied Astronomy, 
  St.~Petersburg

 \bibitem[1999]{petrov99}
  Petrov, L. 1999,
  in Proc. of the 13th Working Meeting on
  European VLBI for Geodesy and Astrometry, 
  ed. by W.~Schl{\"u}ter \& H. Hase, 144--151

 \bibitem[2004]{petrov04}
  Petrov, L. \& Boy, J.-P. 2004,
  J. Geophys. Res., 109(B3), B03405

 \bibitem[2002]{scherneck02}
  Scherneck, H.-G., Haas, R., \& Bos, M. S. 2002,
  in TMR network FMRX-CT96-0071 Scientific Report 1996-2001,
  ed. by J. Campbell, R. Haas \& A. Nothnagel

 \bibitem[2002]{schluter02}
  Schl{\"u}ter, W., Himwich, E. Nothnagel, A., 
  Vandenberg, N. \& Whitney, A. 2002,
   Adv. Space Res. 30(2), 145--150
  
 \bibitem[2004]{schuh04}
  Schuh, H., Boehm, J., Bolotin, S., Capallo, R.,
  Elgered, G., Engelhardt, G., Haas, R., Hanada, H.,
  Hobiger, T., Ichikawa, R., Klioner, S., Ma, C.,
  MacMillan, D., Malkin, Z., Matsusaka, S., Niell, A.,
  Nothnagel, A., Schwegmann, W., Sovers, O.,
  Tesmer, V. \& Titov, O. 2004,
  Report of IVS WG3.6 - Data analysis,
  available at http://ivscc.gsfc.nasa.gov/about/wg/wg3/index.html

 \bibitem[2004]{shapiro04}
  Shapiro, S. S., Davis, J. L., Lebach, D. E. \& Gregory J. S. 2004,
  Phys. Rev. Lett. 92(12), 121101

 \bibitem[1985]{sovers85}
  Sovers, O. J. \& Ma, C. 1985,
  in NASA JPL TDA Progress report 42--83, 101--112

 \bibitem[1994]{sovers94}
  Sovers, O. J. 1994
  Geophys. Res. Lett. 21, 357--360
 
 \bibitem[1996]{sovers96}
  Sovers, O. J. \& Jacobs, C. S. 1996,
  JPL Publication 83--39, Rev. 6

 \bibitem[1998]{sovers98}
  Sovers, O. J. \& Fanselow, J. L., 1998
  Rev. Mod. Phys., 70(4), 1393--1454

 \bibitem[2004]{stoyanov04}
  Stoyanov, B., Haas, R. \& Gradinarsky, L. 2004,
  in IVS 2004 General Meeting Proceedings, 
  ed. by N. R. Vandenberg \& K. D. Baver, 471--475
  
 \bibitem[2004]{tesmer04}
  Tesmer, V. \& Kutterer, H.-J. 2004,
  in IVS 2004 General Meeting Proceedings, 
  ed. by N. R. Vandenberg \& K.~D.~Baver,296--300

 \bibitem[2002]{titov01}
  Titov, O., Tesmer, V. \& B{\"o}hm, J. 2001,
  AUSLIG Technical Report~7

 \bibitem[2004]{titov04}
  Titov, O., Tesmer, V. \& B{\"o}hm, J. 2004,
  in IVS 2004 General Meeting Proceedings, 
  ed. by N. R. Vandenberg \& K.~D.~Baver, 267--271
  
\end{thebibliography}
\end{document}